\documentclass[aps,prb,superscriptaddress,reprint,showpacs,twocolumn,
preprintnumbers,amsmath,amssymb,floatfix]{revtex4-1}
\usepackage{graphicx}
\usepackage{epstopdf}
\usepackage{color}
\usepackage[latin1]{inputenc}
\usepackage{amssymb}
\usepackage{bm}
\usepackage{stmaryrd}
\usepackage{wasysym}
\usepackage{subfigure}

\newcommand{\fnmr}{$^{19}$F }
\newcommand{\tsl}{$1/T_1$ }
\newcommand{\musr}{$\mu$SR }

\begin{document}

\title{Competing effects of Mn and Y doping on the low-energy excitations and phase diagram of La$_{1-y}$Y$_{y}$Fe$_{1-x}$Mn$_x$AsO$_{0.89}$F$_{0.11}$  iron-based superconductors}
\author{M. Moroni}
\email{matteo.moroni01@universitadipavia.it}
\affiliation{Department of Physics, University of Pavia-CNISM,
I-27100 Pavia, Italy}
\author{S. Sanna}
\affiliation{Department of Physics, University of Pavia-CNISM,
I-27100 Pavia, Italy}
\author{G. Lamura}
\affiliation{CNR-SPIN and Universit\'a di Genova, via Dodecaneso
33, I-16146 Genova, Italy}
\author{T. Shiroka}
\affiliation{Laboratorium f\"{u}r Festk\"{o}rperphysik, ETH
H\"{o}nggerberg, CH-8093 Z\"{u}rich, Switzerland} \affiliation{
Paul Scherrer Institut, CH-5232 Villigen PSI, Switzerland }
\author{R. De Renzi}
\affiliation{Department of Physics and Earth Sciences, University
of Parma-CNISM, I-43124 Parma, Italy}
\author{R. Kappenberger} \affiliation{Leibniz-Institut f\"ur
Festk\"orper- und Werkstoffforschung (IFW) Dresden, 01171 Dresden,
Germany}
\author{M.A. Afrassa} \affiliation{Leibniz-Institut f\"ur
Festk\"orper- und Werkstoffforschung (IFW) Dresden, 01171 Dresden,
Germany}\affiliation{Addis Ababa University, College of Natural
Science, Department of Physics, Addis Ababa, Ethiopia}
\author{S. Wurmehl} \affiliation{Leibniz-Institut f\"ur
Festk\"orper- und Werkstoffforschung (IFW) Dresden, 01171 Dresden,
Germany} \affiliation{Institute for Solid State Physics, Dresden
Technical University, TU-Dresden, 01062 Dresden, Germany}
\author{A.U.B. Wolter} \affiliation{Leibniz-Institut f\"ur
Festk\"orper- und Werkstoffforschung (IFW) Dresden, 01171 Dresden,
Germany}
\author{B. B\"uchner} \affiliation{Leibniz-Institut f\"ur
Festk\"orper- und Werkstoffforschung (IFW) Dresden, 01171 Dresden,
Germany} \affiliation{Institute for Solid State Physics, Dresden
Technical University, TU-Dresden, 01062 Dresden, Germany}
\author{P. Carretta}
\affiliation{Department of Physics, University of Pavia-CNISM,
I-27100 Pavia, Italy}

\begin{abstract}
Muon Spin Rotation (\musr) and \fnmr Nuclear Magnetic Resonance
(NMR) measurements were performed to investigate the effect of Mn
for Fe substitutions in
La$_{1-y}$Y$_{y}$Fe$_{1-x}$Mn$_x$AsO$_{0.89}$F$_{0.11}$
superconductors. While for $y = 0$ a very low critical
concentration of Mn ($x = 0.2$\%) is needed to quench
superconductivity, as $y$ increases the negative chemical pressure
introduced by Y for La substitution stabilizes superconductivity
and for $y= 20$\% it is suppressed at Mn contents an order of
magnitude larger. A magnetic phase arises once superconductivity
is suppressed both for $y$=0 and for $y= 20$\%. Low-energy spin
fluctuations give rise to a peak in \fnmr NMR $1/T_1$
with an onset well above the superconducting transition temperature
and whose magnitude increases with $x$. Also the static magnetic
correlations probed by \fnmr NMR linewidth measurements show a
marked increase with Mn content. The disruption of
superconductivity and the onset of the magnetic ground-state are
discussed in the light of the proximity of
LaFeAsO$_{0.89}$F$_{0.11}$ to a quantum critical point.
\end{abstract}


\maketitle


\section{\label{sec:intro}Introduction}

The introduction of impurities in superconductors is a well known
approach to probe the local response function and to unravel their
intrinsic microscopic properties.\cite{Alloul} Both spinless and
paramagnetic impurities perturb the local electronic environment
and cause a significant change in the spin polarization around
them. When the spin correlations are particularly enhanced, as it
is the case in the proximity of a quantum critical point
(QCP),\cite{QCP,Sachdev} or when the amount of impurities starts
to be significant, cooperative effects become relevant and marked
changes in the superconducting transition temperature are
observed, eventually leading to the appearance of a magnetic
order.\cite{Alloul1}

In the pnictides extensive studies on the effect of impurities on
the superconducting ground-state have been reported
\cite{Sato1,Lee,Sato1B,SannaPRL,Sanna1su4,Bobroff} and the most
dramatic and yet not fully understood effect is induced by Mn for
Fe substitution in the optimally electron-doped
LaFeAsO$_{0.89}$F$_{0.11}$.\cite{Hammerath2014} In this material
it is sufficient to introduce a tiny amount of Mn, as low as
0.2\%, to fully quench superconductivity. It has been shown that
at this doping level there is a divergence of the in-plane
correlation length, characteristic of a two-dimensional (2D)
antiferromagnetically correlated metal approaching a quantum
critical point.\cite{Hammerath2014} This QCP separates the
superconducting phase from a magnetic ground-state developing at
Mn contents above 0.2\%. Originally it was suggested that Mn
impurities could lead to a shift in the spectral weight of the
fluctuations from $(0,\pi)$ (stripe wave-vector) to $(\pi,\pi)$
(N\'eel wave-vector) \cite{Tucker} and accordingly to a
suppression of interband pairing processes.\cite{Millis} However,
no evidence of a N\'eel phase in Mn-doped
LaFeAsO$_{0.89}$F$_{0.11}$ has been ever reported and recent
experiments seem rather to suggest that the magnetic order is
still characterized by a stripe collinear
arrangement.\cite{Moroni2016} It is interesting to notice that
such a marked effect is observed for LaFeAsO$_{0.89}$F$_{0.11}$
only, whereas LnFeAsO$_{0.89}$F$_{0.11}$ with smaller lanthanide
ions (e.g. for Ln=Sm) shows a much less dramatic effect and much
larger amounts of Mn are needed to suppress superconductivity.
\cite{sato2010,singh2013}

In this paper we present a study of the effect of Mn doping in
LaFeAsO$_{0.89}$F$_{0.11}$ where La is partially substituted by Y,
for doping levels up to 20 \%. By combining muon spin rotation
$\mu$SR with superconducting quantum interference device (SQUID)
magnetometry we were able to draw the phase diagram of
La$_{1-y}$Y$_{y}$Fe$_{1-x}$Mn$_x$AsO$_{0.89}$F$_{0.11}$, at fixed
Y content as a function of the Mn doping level and at fixed Mn
doping as a function of the Y doping level. It is shown that Y
doping causes a significant shift of the QCP observed in the $y=0$
system and that magnetism arises only for $x> 5$ \%, for $y= 20$
\%. $^{19}$F nuclear spin-lattice relaxation measurements evidence
the enhancement of low-frequency dynamics already present in the
normal phase of the samples without Mn. The mechanism giving rise
to the onset of the magnetic phase and the suppression of
superconductivity are discussed in the light of recent theoretical
models.

\section{\label{sec:exp_res}Experimental methods and results}

Two series of polycrystalline
La$_{1-y}$\-Y$_{y}$\-Fe$_{1-x}$\-Mn$_x$\-As\-O$_{0.89}$\-F$_{0.11}$
samples have been studied: the first one with fixed $y=20$\%
yttrium content (LaY20 hereafter) and nominal Mn contents ranging
from $x=0$\% to $20$\%, while the second one was prepared with
fixed $x=0.5$\% Mn content and $y=0.5$\%, $1$\%, $5$\%, $10$\%,
$20$\%, $23$\% yttrium contents (LaYMn05 hereafter). The samples
were synthesized using a two-step solid-state
reaction~\cite{Alfonsov2011}. Details on sample preparation and
characterization by means of powder x-ray diffraction, electron
microscopy (SEM) and SQUID magnetometry, used to determine $T_c$,
have been already partially reported in
Refs.~\onlinecite{SupMoroni2015,Sabine2016}. Electron microscopy
WDX revealed that Y and Mn contents are quite close to the nominal
ones. All the samples are optimally electron doped with a nominal
fluorine content of 11\%. The results obtained in the LaY20 series
will be compared to those already derived for
LaFe$_{1-x}$Mn$_{x}$AsO$_{0.89}$F$_{0.11}$ (LaY0
hereafter).\cite{Hammerath2014} It is pointed out that the LaY0
series \cite{Sato1} was not grown with exactly the same procedure
as the LaY20 series. Although this may lead to slight changes in
the phase diagram this will not affect the analysis and the
conclusions presented in this work.

\begin{figure}[h!]
\includegraphics[height=5.5cm,
keepaspectratio]{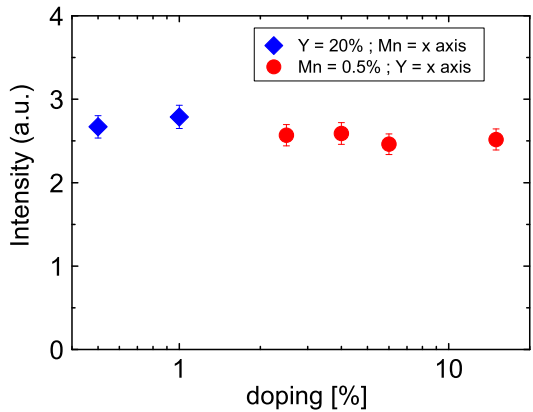} \caption{(Color online) Intensity
of $^{19}$F NMR signal at 1.5 Tesla, normalized by the sample
mass, at room temperature for selected LaY20 and LaYMn05 samples
(see legend).} \label{fluorine}
\end{figure}

The intensity of the \fnmr NMR signal was measured at room
temperature in order to check the effective fluorine content both
for the LaY20 and for the LaYMn05  series. The results, reported
in Fig.~\ref{fluorine} show that the absolute fluorine
stoichiometry is constant in each sample series within $\pm
0.5\%$.

\subsection{\label{sec:musr}Muon spin relaxation results}
\begin{figure*}[t]
\includegraphics[height=6.6cm,
keepaspectratio]{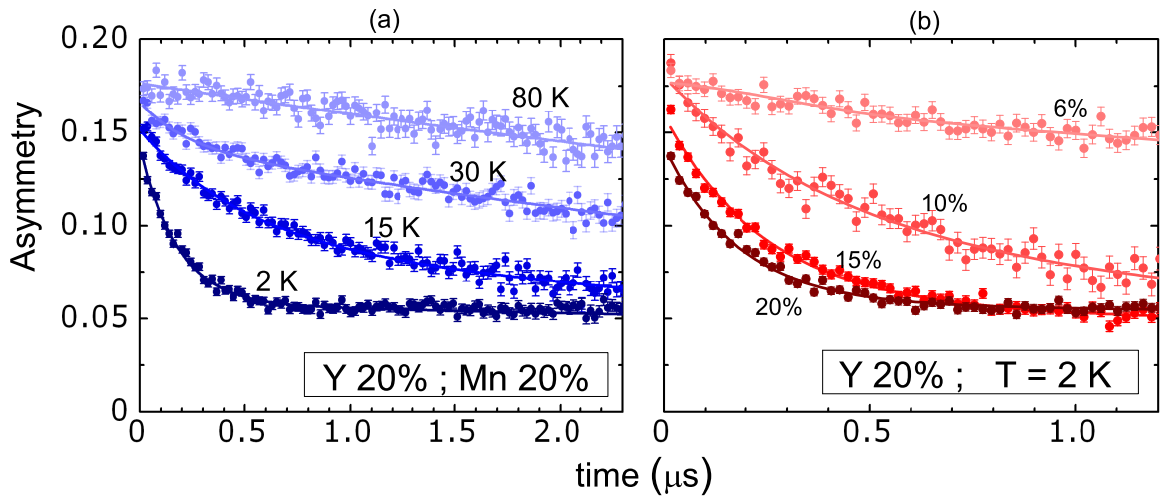} %
\caption{(Color online) (a) Zero field $\mu$SR signal for the
La$_{0.8}$Y$_{0.2}$Fe$_{1-x}$Mn$_{x}$AsO$_{0.89}$F$_{0.11}$ sample
with $x=20$\%, measured at different temperatures. (b) ZF $\mu$SR
signal for LaY20 samples with $x=$6\%,10\%,15\% and 20\%, at $T=
2$ K. Solid lines in (a) and (b) represent the best fits to
Eq.~\ref{asymfit}.} \label{asym}
\end{figure*}
In a muon spin relaxation ($\mu$SR) experiment 100\%
spin-polarized positive muons ($\mu^+$) are implanted uniformly
into the sample. The $S_{\mu}= 1/2$ muon spin acts as a magnetic
probe, precessing around the local magnetic  field  $B_\mu$ at a
frequency $\nu= \gamma_\mu B_\mu/2\pi$, where $\gamma_\mu=
2\pi\times 135.53 $ MHz/T is the muon gyromagnetic ratio. When the
muons decay they emit a positron preferentially along the
direction of their spins. Hence, by counting the positrons emitted
along a given direction one can reconstruct the time dependence of
the muon decay asymmetry $A(t)$, proportional to the time
evolution of the muon spin
polarization.\cite{DrewBlundell2008,YaouancBook}

In order to probe the local magnetic properties of
La$_{1-y}$Y$_{y}$Fe$_{1-x}$Mn$_x$AsO$_{0.89}$F$_{0.11}$, zero
field (ZF) and longitudinal field (LF) measurements were carried
out at the Paul Sherrer Institut (PSI) with the Dolly instrument
of S$\mu$S facility. ZF measurements are extremely sensitive to
spontaneous magnetism since in this configuration the local field
at the muon site originates from the internal magnetic order only.
On the other hand, LF measurements represent a useful tool to
study the spin dynamics and can conveniently be used to
distinguish between static and dynamic
magnetism~\cite{DrewBlundell2008,YaouancBook}.

Figure~\ref{asym} shows the typical time dependence of the ZF
$\mu$SR asymmetry at different temperatures
for the samples that display a magnetic order below $T_\mathrm{N}$.
The time evolution of the muon asymmetry could be fit with the
following standard function:
\begin{equation} \label{asymfit}
A(t)= A_0\left[ f_\parallel e^{-\lambda_\parallel t} + f_\perp
G(t, B_{\mu}) \right]\mbox{ ,}
\end{equation}
where $A_0$ is the initial $\mu$SR asymmetry, while $f_\parallel$
and $f_\perp$ are the longitudinal
($\textbf{B}_\mu\parallel\textbf{S}_\mu$) and transverse
($\textbf{B}_\mu\perp\textbf{S}_\mu$) fractions of the asymmetry,
respectively. The function $G(t, B_{\mu})= \exp(-\lambda_\perp t)$
determines the time dependence of the transverse component,
whereas the longitudinal one decays exponentially with a decay
rate $\lambda_\parallel$.

\begin{figure}[t]
\includegraphics[height=9.5cm,
keepaspectratio]{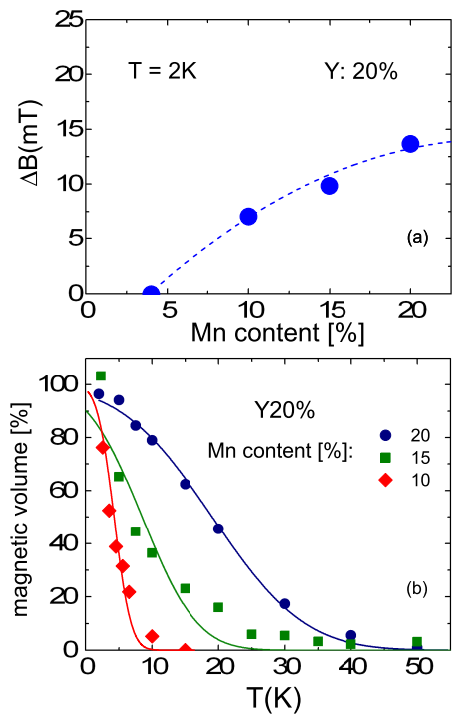} %
\caption{(Color online) (a) The values of $\Delta B $ at 2 K, for
all the samples showing a magnetic order (see Fig. 7), obtained
from the fit of the muon asymmetry to Eq.~\ref{asymfit}. The
dashed line is a guide to the eye. (b) The magnetic volume
fraction temperature dependence is shown for $y=20\%$ and
$x=10\%$, $15\%$ and $20\%$. The solid lines are best fits to
Eq.~\ref{error}.} \label{internal_field}
\end{figure}

At high temperature ($T>$ 30 K) the samples of the LaY20 and
LaYMn05 series are in the paramagnetic regime and the muon
asymmetry can be fit by setting $f_\perp$=0, with decay rates
$\lambda_\parallel\sim 0.09$ $\mu$s$^{-1}$. Upon decreasing the
temperature  a fast decaying component $f_\perp$ emerges in the
LaY20 samples with $x\geq10$\%, evidencing the presence of
overdamped oscillations in the muon asymmetry. A similar behavior
is observed for samples close to the magnetic superconducting
boundary \cite{carretta2013,prando2013} and reflects the presence
of a significant distribution of local magnetic fields, typically
observed when a short range AF magnetic order
develops.\cite{carretta2013} The size of the internal fields is of
the order of the field distribution $\Delta B_{\mu}$, which can be
roughly estimated as $\Delta B_{\mu} = \lambda_\perp
/\gamma_{\mu}$. The values of $\Delta B_{\mu} $ obtained from the
fit of the data with Eq.~\ref{asymfit}, of the order of 10 mT, are
shown in Fig~\ref{internal_field}(a). The static character of the
magnetism developing at $T< T_\mathrm{N}$ has been confirmed by LF
$\mu$SR experiment which have shown that a field of about 100 mT
is enough to completely recover the initial muon asymmetry at 2 K.
At variance, all LaY20 samples with $x< 6$\% and all the samples
of the LYaMn05 series do not display a spontaneous magnetic order
down to 2 K.

The sample magnetic volume fraction $V_m$, namely the fraction of
the sample volume where the muons sense the magnetic order, can be
derived from $f_\parallel$. From simple geometric arguments
\cite{DrewBlundell2008,YaouancBook} it can be shown that  in a
polycrystalline sample with 100\% magnetic volume fraction
$f_\parallel=1/3$ and that in general one can write $V_m(T)=
3/2(1-f_\parallel(T))$. The temperature dependence of $V_m$
(Fig.~\ref{internal_field}b) shows that the full magnetic volume
condition is reached only at low temperatures for all the
magnetically ordered samples (LaY20 with $x\geq10\%$). The
magnetic ordering temperature can be estimated by fitting $V_m(T)$
to the error function
\begin{equation} \label{error}
V_m(T)=\frac{1}{2}\left(
1-\mbox{erf}\left(\frac{T-T_\mathrm{N}}{\sqrt{2}\Delta
T_\mathrm{N}}\right) \right)
\end{equation}
which assumes the presence of a Gaussian distribution of local
transition temperatures centered around the average value
$T_\mathrm{N}$. The results are reported in the phase diagram in
Fig.~\ref{phase}.

\subsection{\label{sec:nmr}Nuclear magnetic resonance results}
$^{19}$F NMR experiments were performed on LaY20 samples in order
to complete the study reported in Ref~\onlinecite{Moroni2015}. The
Y for La substitution results in a system with higher chemical
pressure  with respect to La1111 (La$^{3+}$ and Y$^{3+}$ ionic
radii are 103 pm and 90 pm, respectively), without introducing
paramagnetic lanthanide ions, such as Sm$^{3+}$, which would
significantly affect the $^{19}$F spin-lattice relaxation rate
(1/T$_1$). \cite{Prando2010}

The polycrystalline samples were milled to fine powders in order
to improve the radiofrequency penetration. All the measurements
were performed in a magnetic field of 1.36 T, in the temperature
range between 4 K and 100 K. For a few selected samples the
temperature range was extended up to 200 K to precisely estimate
the high temperature \fnmr\tsl trend.

\begin{figure}[h]
\includegraphics[height=6cm,
keepaspectratio]{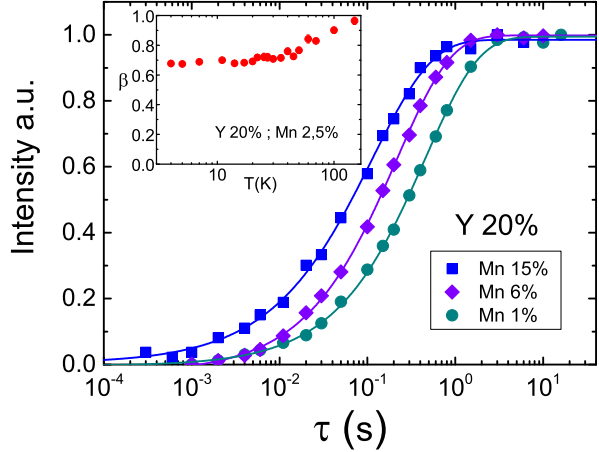} \caption{(Color online)
$^{19}$F nuclear magnetization recovery for
La$_{0.8}$Y$_{0.2}$Fe$_{1-x}$Mn$_{x}$AsO$_{0.89}$F$_{0.11}$ at
22K, around the \tsl peak, for different values of $x$ (see
legend). The solid lines are fits to Eq.~\ref{recovery}. Inset:
temperature dependence of the stretching exponent $\beta$ used to
fit the longitudinal nuclear magnetization recovery for $x= 2.5$\%
and $y=20$\%.  } \label{recoveries}
\end{figure}

The \fnmr spin lattice relaxation rate was estimated by fitting
the recovery of the longitudinal magnetization $M_z(\tau)$ after a
saturation recovery pulse sequence \mbox{($\frac{\pi}{2} -
\tau-\frac{\pi}{2}-\tau_{echo}-\pi$)}. For all the samples the
recovery could be nicely fit to a stretched exponential (see
Fig.~\ref{recoveries}):
\begin{equation} \label{recovery}
M_z(\tau)=M_0[1- e^{-(\tau/T_1)^\beta}]\mbox{ ,}
\end{equation}
with $M_0$ the nuclear magnetization at thermal equilibrium and
$\beta$ the stretching exponent.
\begin{figure}[h]
\includegraphics[height=7cm,
keepaspectratio]{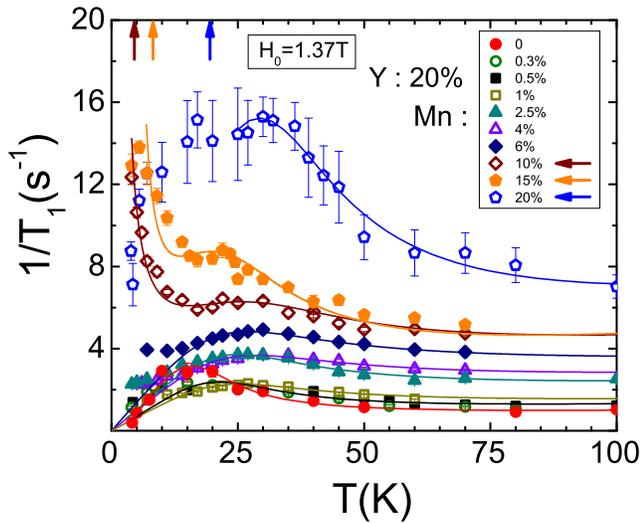} \caption{(Color online) Temperature
dependence of $^{19}$F NMR 1/T$_1$ for
La$_{0.8}$Y$_{0.2}$Fe$_{1-x}$Mn$_{x}$AsO$_{0.89}$F$_{0.11}$ for Mn
doping levels up to $x=$20\%. The solid lines are fits of the data
according to Eq.~\ref{T1eq} in the text. The arrows indicate
$T_\mathrm{N}$ for the magnetic samples (from left to right:
$x=10\%$, $x=15\%$, $x=20\%$)} \label{t1}
\end{figure}
The stretching exponent progressively decreased on cooling below
100 K and it was found in the range $0.5\leqslant \beta \leqslant 1$
for all samples (see the inset to Fig.~\ref{recoveries}). This
behaviour indicates the presence of a distribution of spin lattice
relaxation times which is a common feature of disordered systems
and in our case it is probably due to the different inequivalent
impurity configurations resulting from  Y and Mn doping. In fact,
the low temperature values of $\beta$ get smaller on increasing
the Mn content, namely the number of impurities.

The temperature dependence of \tsl in LaY20, for Mn contents
ranging from $x=0\%$ up to $x=20\%$, is shown in Fig.~\ref{t1}.
While at high temperature \tsl displays a linear Korringa behavior
(see Ref. \onlinecite{Moroni2015}) typical of weakly correlated
metals, below 80K the spin lattice relaxation rate progressively
increases on cooling, giving rise to a broad peak around 25K. It
is remarked that this increase starts well above T$_c$ or well
above T$_N$, for the magnetically ordered samples. Insights on the
nature of the peak can be gained by observing its evolution upon
changing the magnitude of the external magnetic field $\vec H_0$.
Measurements in a lower field of 0.75 T revealed that while at
high temperature \tsl  is only weakly field dependent, the
magnitude of the peak at 25 K is significantly
enhanced,\cite{Moroni2015} which is exactly the behaviour expected
for slow dynamics with a characteristic frequency in the MHz
range, close to the Larmor frequency $\omega_0$.

The behavior of \tsl below 25~K depends on the Mn doping level: in
samples with Mn doping below 10\% the spin lattice relaxation rate
decreases with lowering temperature, while for samples with higher
Mn doping we observed a steep increase of \tsl with a divergence
at temperatures approaching the magnetic transition temperature
determined by $\mu$SR. This behavior is associated with the
critical divergence of the spin correlation length on approaching
the magnetic transition, which yields a power law divergence of
$1/T_1\propto (T-T_\mathrm{N})^{-\alpha}$.

\begin{figure}[h]
\includegraphics[height=6.2cm,
keepaspectratio]{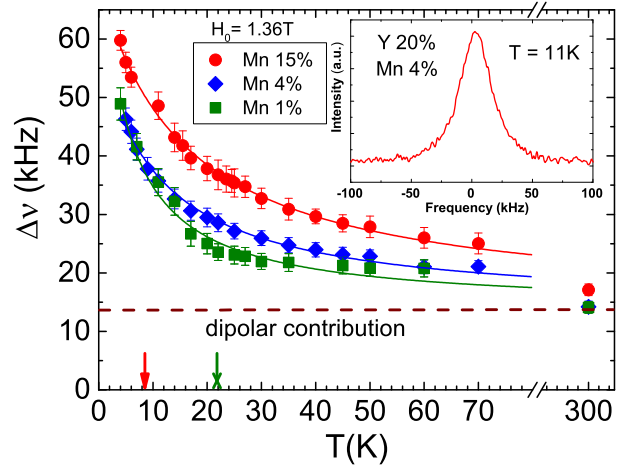} \caption{(Color online) \fnmr NMR full
width at half intensity for three representative samples of the
LaY20 series. Solid lines are best fits according to a Curie-Weiss
law while the arrows indicate $T_c$ of the $x=1$\% sample (green
arrow) and $T_\mathrm{N}$ of the $x=15$\% sample (red arrow).
Inset: a typical \fnmr NMR spectrum ($x=4$\%, $y=20$\%, $T=11$ K).
} \label{FWHM}
\end{figure}

Further insights on the effects of Mn doping can be gained from
the study of the temperature dependence of the \fnmr NMR linewidth
$\Delta\nu$, directly related to the amplitude of the staggered
magnetization developing around the Mn
impurity\cite{linewidth_parigi}. $\Delta\nu$ was derived from the
Fast Fourier Transform of half of the echo signal after a Hahn
spin-echo pulse sequence.  As it can be seen in Fig.~\ref{FWHM} by
increasing the Mn content a marked increase of $\Delta\nu$ is
observed. The data reported were fitted with a Curie-Weiss law
$\Delta\nu= (\Delta\nu )_0+ C/(T+\Theta)$ (see solid lines in
Fig.~\ref{FWHM}). The temperature independent term
$\Delta\nu_0\sim 14$ kHz estimated from the fit of the data up to
$T= 300$ K, is in very good agreement with the value 13.5 kHz
estimated for the nuclear dipole-dipole interaction derived from
lattice sums. About 80\% of the second moment is due to F-La
nuclear dipole interaction and about 19.5\% to F-F interaction,
while only a minor contribution arises from F-As interactions.
This term practically does not change by increasing the Mn doping
since the lattice parameters change by less than 1.2\% between
$x=0$ and $x=20$\% \cite{SupMoroni2015} and the dipolar
contribution of $^{55}$Mn nuclei for $x= 20$\% would cause a
change by less than 1\% of the linewidth. The results of the fits
are summarized in Table.~\ref{tab:curie}.

\begin{table}[ht]
\caption{Curie constant $C$ and Curie-Weiss temperature $\theta$
obtained from the analysis of the temperature evolution of the
$^{19}$F NMR line width $\Delta\nu$ shown in Fig.~\ref{FWHM} for
La$_{0.8}$Y$_{0.2}$Fe$_{1-x}$Mn$_{x}$AsO$_{0.89}$F$_{0.11}$.}
\label{tab:curie}
\begin{ruledtabular}
\begin{tabular}{||p{8mm}|  l | p{8mm}||}
Mn (\%) & $C$ (kHz$\cdot$K) & $\theta$ (K) \\
\hline
1 & $300\pm30$ & $4\pm1$  \\
4 & $490\pm20$ & $11\pm1$  \\
15 & $870\pm20$ & $16\pm1$  \\
\end{tabular}
\end{ruledtabular}
\end{table}

\begin{figure}[h!]
\includegraphics[height=6.5cm,
keepaspectratio]{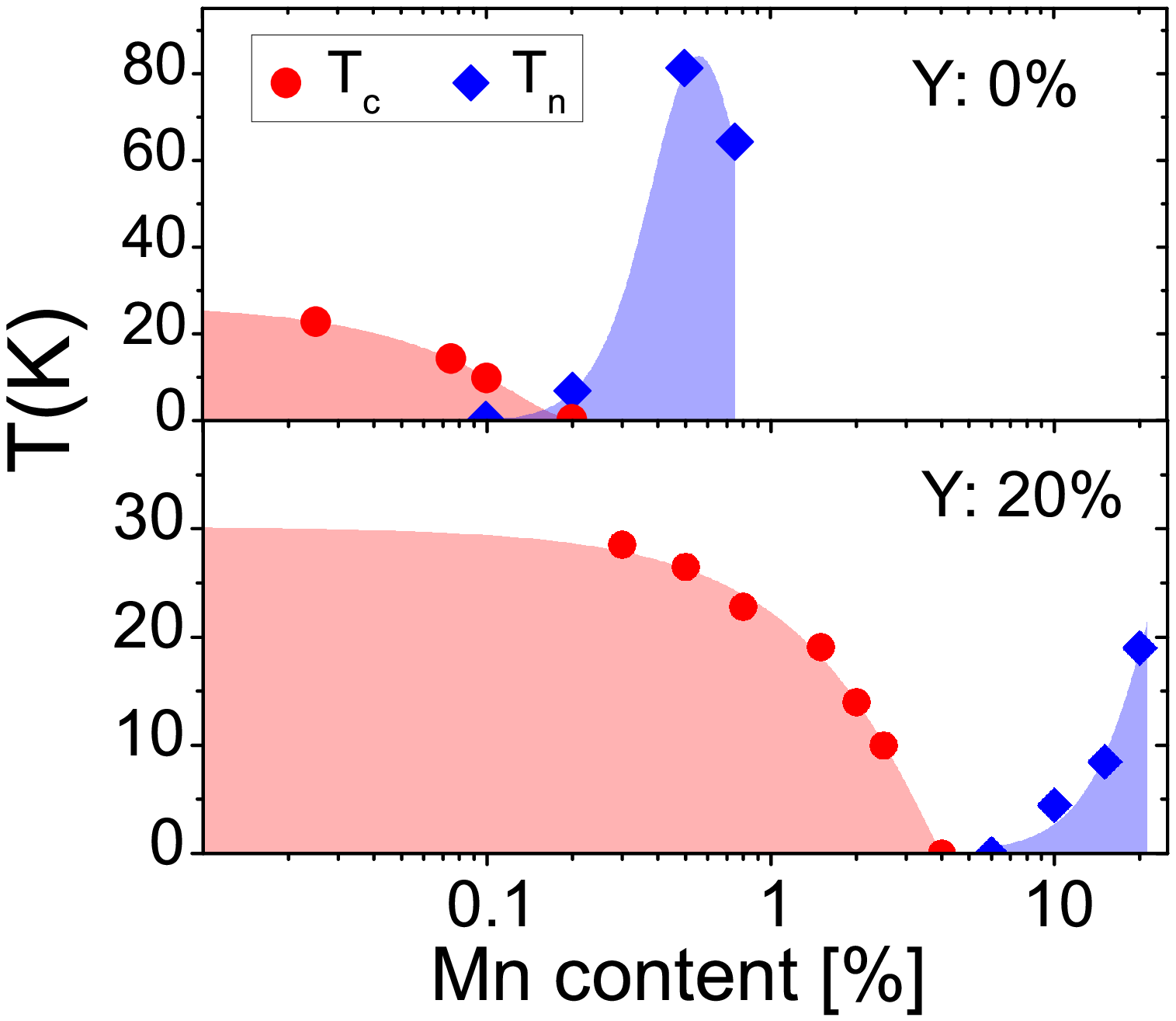} \caption{(Color online)
Phase diagram of
La\-Fe$_{1-x}$\-Mn$_{x}$\-As\-O$_{0.89}$\-F$_{0.11}$ (top) and of
La$_{0.8}$\-Y$_{0.2}$\-Fe$_{1-x}$\-Mn$_{x}$\-As\-O$_{0.89}$\-F$_{0.11}$
(bottom). The red and blue shaded areas are the superconductive
and the magneitc phases, respectively. The magnetic transition
temperature (blue diamonds) was determined by ZF-\musr while the
supercunducting transition temperature $T_c$ (red circles) was
determined from SQUID magnetization measurements} \label{phase}
\end{figure}

\begin{figure}[h!]
\includegraphics[height=5.3cm,
keepaspectratio]{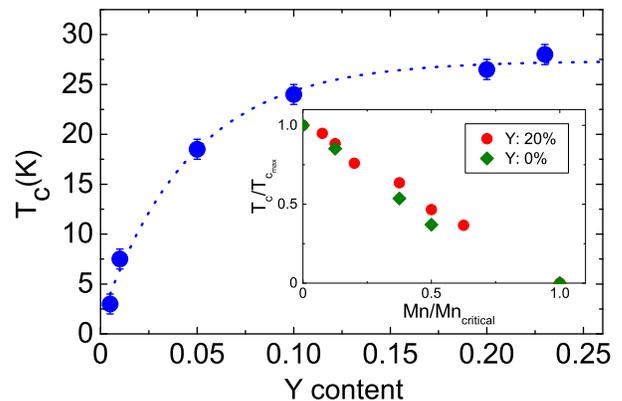} \caption{(Color online) Critical
temperatures for the LaYMn05 samples studied in this work. The
dashed line is a guide to the eye. Inset: $T_c(x)/T_c(x=0)$ versus
the Mn content normalized by the critical content causing the
vanishing of $T_c$ for LaY20 and LaY0
(Ref.~\onlinecite{Hammerath2014}) series (see text).}
\label{phaseY}
\end{figure}

\section{\label{sec:discussion}Discussion}

The phase diagram of LaY20, and for comparison that of LaY0
\cite{Hammerath2014}, derived from SQUID magnetization and $\mu$SR
measurements, are shown in Fig.~\ref{phase} as a function of Mn
content. These data show that the substitution of La with Y causes
a sizeable increase of the critical Mn threshold $x_c$ required to
suppress superconductivity, with an increase from 0.2\% to 4\% on
increasing the Y content from 0 to 20\%. In the latter system the
threshold is comparable to the one observed \cite{sato2010,
singh2013} when smaller paramagnetic ions fully substitute La,
namely one has  $x_c$= 4 and 8\% for \emph{Ln}=Nd and Sm,
respectively. The results on LaY20, where La is partially
substituted by the smaller but non-magnetic Y ion, clearly
evidence that the electronic properties of the
LaFeAsO$_{0.89}$F$_{0.11}$ are significantly affected by the
chemical pressure or strain induced by the different radii of the
lanthanide ions on the FeAs planes. The effect of the chemical
pressure is further evidenced in Fig.~\ref{phaseY} showing $T_c$
as a function of the La/Y substitution in
La$_{1-y}$Y$_{y}$Fe$_{0.995}$Mn$_{0.005}$AsO$_{0.89}$F$_{0.11}$.
Superconductivity, suppressed by the tiny quantity of 0.5\% of Mn
for the end $y= 0$ member, is gradually restored by increasing the
Y content or, in other terms, the chemical pressure.

When superconductivity is fully suppressed a magnetic order arises
both in the LaY0 and in the LaY20 series, as shown in
Fig.~\ref{phase}. This behavior suggests that the two orders are
competing and that for LaY0 a quantum critical point is separating
the superconducting and magnetic phases, as supported by the
previous\cite{Hammerath2014} analysis of the temperature
dependence of the magnetic correlation length. On the other hand,
it should be pointed out that for LaY20 a crossover region where
both T$_c$ and T$_N$ are zero is observed for 4\% $< x<$ 6\%.

The onset of a magnetic order for just 0.2\% of Mn in LaY0
indicates that in this compound significant electronic
correlations must be present. If the ratio between Hubbard
repulsion and the hopping integral associated with the $i$-th band
$U/t_i$ is sizeable, a significant enhancement of the local spin
susceptibility occurs.\cite{Brian} In the iron-based
superconductors quite different behaviours may be observed for the
electrons in the five bands crossing the Fermi level and Hund's
coupling may even lead to orbital selective Mott
transitions.\cite{Capone} However, for simplicity in the following
discussion we will consider that in LaFeAsO$_{0.89}$F$_{0.11}$ the
behaviour can be described by an average value of $U/t$. If $U/t$
is close to a critical value leading to charge localization, the
amplitude and the extension of the spin polarization around the Mn
impurity significantly increase with respect to a weakly
correlated metal\cite{Brian} and even a tiny amount of impurities
may drive the system towards a magnetic ground-state. Hence, the
undoped LaFeAsO$_{0.89}$F$_{0.11}$ superconductor must be very
close to a QCP since a significant change in the electronic
properties occurs by perturbing the system with tiny Mn amounts.
This aspect is further supported by the charge localization
observed in the LaY0 for Mn contents above $x_c$ and by the
significant changes in the $c$ axis lattice parameter.\cite{Sato1}
Moreover, as we have previously mentioned, we found that the
behaviour of the spin correlation length is that expected for a
two-dimensional antiferromagnet close to a QCP.
\cite{Hammerath2014} Hence the quenching of superconductivity
should not be ascribed to a pair breaking effect, where the
suppression of superconductivity yields the recovery of the normal
metallic state, but to a quantum phase transition affecting the
LaY0 electronic ground-state.

The increase in the chemical pressure induced by Y doping causes
an increase in the metallic character and a decrease in $U/t$.
Accordingly, the spin polarization around the Mn impurity is
reduced and larger Mn contents are needed to induce a magnetic
order which, in any case, appears to be characterized by an order
parameter which is weaker than that observed in the LaY0 system
(Fig.~\ref{phase}). However, the behaviour of the LaY0 and LaY20
series becomes similar (see the inset of Fig.~\ref{phaseY}) once
the phase diagram is rescaled by the critical Mn content $x_c$ and
the superconducting transition temperature for T$_c(x=0)$. Namely
the Mn doping level  $x_c$ yielding the quantum critical point is
renormalized by the decrease in $U/t$.

\begin{figure}[h!]
\includegraphics[height=8cm,
keepaspectratio]{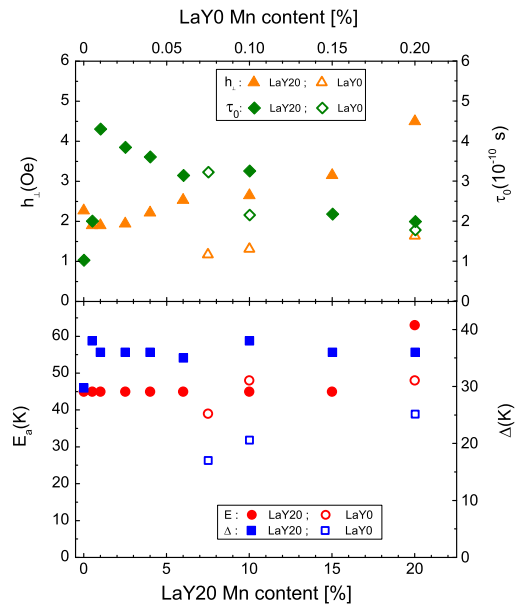}  \caption{(Color online)
Parameters extracted from the fit of $1/T_1$ to Eq. 6 as a
function of the Mn content for the LaY20 series (filled symbols,
bottom horizontal axes) and for the LaY0 series (open symbols, top
horizontal axes). Top panel: mean value of the local fluctuating
magnetic field $h_\perp$ (left) and correlation time $\tau_0$
(right). Bottom panel: energy barrier $E_a$ (left) and width of
the energy barrier distribution $\Delta$ (right).  }
\label{fit_parameters}
\end{figure}

Further insights on the effect of Mn in LaY0 and LaY20 can be
derived from the analysis of the $^{19}$F NMR spin-lattice
relaxation rate. As it is shown in Fig.~\ref{t1} a broad peak in
$1/T_1$, which is increasing with the Mn content, is detected
around 25 K. That peak, observed both in the LaY0 and in the LaY20
series, is a very general feature of these systems since it is
present both in superconducting and magnetic samples and also in
the sample without Mn doping,\cite{Hammerath2013,Moroni2015}
although slightly shifted to lower temperatures. The introduction
of increasing Mn contents gives rise to a progressive enhancement
of the peak magnitude, suggesting that the presence of
paramagnetic impurities strengthens the low-frequency dynamics
already present in the pure compound. Its dependence on the
magnetic field intensity \cite{Moroni2015} indicates that it has
to be associated with low-frequency fluctuations (MHz range). This
peak should not be ascribed to the slowing down of the critical
fluctuations on approaching $T_\mathrm{N}$, which are present only
in the magnetic samples and give rise to a steeper increase in
$1/T_1$ only at lower temperatures (Fig.~\ref{t1}).

The approach devised by Bloembergen-Purcell-Pound\cite{BPPtheory}
(BPP model) is often suited to describe \tsl in presence of
hyperfine field $\vec h(t)$ fluctuations approaching the Larmor
frequency $\omega_0$, namely in the MHz range. The model assumes
that the autocorrelation function for the field fluctuations
decays exponentially:
\begin{equation} \label{autocor}
\langle h_\perp (t+\tau) h_\perp(t)\rangle =\langle h_\perp^2
\rangle e^{-t/\tau_c},
\end{equation}
where $\tau_c$ is the characteristic time of the fluctuations and
$h_\perp$ is the component of the local fluctuating hyperfine
field perpendicular to $\vec{H_0}$, with $\langle
h_\perp^2\rangle$ its mean square amplitude. The spin lattice
relaxation rate, which probes the spectral density at $\omega_0$,
then takes the form:
\begin{equation} \label{eqbpp}
\dfrac{1}{T_1}= \gamma^2 \langle h^2_\perp\rangle
\dfrac{\tau_c(T)}{1+\omega^2_0\tau_c(T)^2}\mbox{ ,}
\end{equation}
where $\gamma$ is the nuclear gyromagnetic ratio. In many
disordered systems, including cuprates,\cite{MHJ2} $\tau_c(T)$ is
described by a thermally activated law $\tau_c(T)= \tau_0\exp
(E_a/T)$, where $E_a$ is the energy barrier and $\tau_0$ the
correlation time at infinite temperature. However, monodispersive
fluctuations cannot explain the broad peaks observed in Mn doped
compounds. A much better result can be obtained by considering a
distribution of energy barriers, and thus of correlation times,
associated with the irregular distribution of Mn impurities.

For simplicity, the energy barrier distribution was taken as
squared, centered around $E_a$ and with a width $\Delta$.
Accordingly Eq.\ref{eqbpp} takes the form:\cite{squareddist}
\begin{equation}\label{T1eq}
\begin{split}
\dfrac{1}{T_1}=\; & \frac{\gamma^2 <h^2_\perp> T}{2\omega_0
\Delta}\biggl[ \arctan\biggl( \omega_0\tau_0
e^{(<E_a>+\Delta)/T}\biggr) \\ & -\arctan\biggl( \omega_0\tau_0
e^{(<E_a>-\Delta)/T}\biggr) \biggr] + c T
\end{split}
\end{equation}
where a linear Korringa-like term $c T$ was added to account for
the high temperature behaviour. Eq.~\ref{T1eq} was used to fit the
\tsl data for all samples with Mn contents lower than 8\%, while
for samples with higher Mn doping a term proportional to
$(T-T_\mathrm{N})^{-\alpha}$ was added to account for the
divergence of \tsl at the magnetic phase transition. As it is
shown in Fig.~\ref{t1} the $1/T_1$ data can be suitably fit to Eq.
6, with the parameters reported in Fig.~\ref{fit_parameters}. The
critical exponent was found to be $\alpha \simeq 1$ both for $x=
10$\% and for $x= 15$\%. Since in quasi-2D antiferromagnets
$1/T_1\sim \xi^z$,\cite{carrettaZn} with $\xi\propto
(T-T_\mathrm{N})^{-\nu}$ the spin correlation length and $z$ and
$\nu$ scaling exponents close to the unity \cite{Boucher}, the
value derived for $\alpha$ appears to be quite reasonable.

The fit parameters shown in Fig.~\ref{fit_parameters} evidence
that the mean value of the energy barrier $E_a$ is nearly constant
as a function of Mn and that the variation of the correlation time
$\tau_0$ of the spin fluctuations is small, in the range of
0.1-0.4 ns. In the LaY0 series $\Delta$ increases with $x$
suggesting that the Mn leads to a distribution of activation
energies which reflects a strong inhomogeneous electronic
environment, even at very small Mn doping levels. For the LaY20
system this distribution is nearly constant and affected by the
disorder induced by the large amount of Y introduced in the
system. The most significant change is the increase in the
amplitude of the local fluctuating field $(<h_\perp^2>)^{1/2}$
with $x$, which indicates that the strength of the local spin
susceptibility in the FeAs plane becomes progressively enhanced by
Mn doping. The enhancement of the local spin susceptibility is
further supported by the analysis of the temperature dependence of
the $^{19}$F NMR line width (Fig.~\ref{FWHM}), which is directly
related to the amplitude of the staggered magnetization developing
around the impurity. The results, summarized in
Table.~\ref{tab:curie}, show that both the Curie constant and the
Curie-Weiss temperature increase as a function of Mn, indicating
that the insertion of Mn strengthens the spin correlations.

The origin of the low-energy fluctuations giving rise to the peak
in $^{19}F$ NMR $1/T_1$ is not yet clear. They seem to be
intrinsic to the system since they are detected also for the LaY20
compound without manganese. Furthermore the related activation
energies $E_a$ and correlation time $\tau_0$ are almost
insensitive to the Mn content indicating that the low-frequency
dynamics is nearly unaltered when approaching the disruption of
superconductivity. Bumps in the 1/$T_1$ vs $T$ behaviour have also
been detected in other optimally electron-doped iron-based
superconductors \cite{Hammerath2013, bossoni2013,
dioguardi2015,bossoni2015} in the same T range where the peak in
$^{19}$F NMR 1/$T_1$ arises in Fig.~5. They have been tentatively
associated with nematic fluctuations \cite{dioguardi2015} or with
the motion of domain walls separating nematic phases
\cite{bossoni2015}. In this scenario the energy barrier $E_a$ may
be related to the one separating the degenerate nematic phases
\cite{fernandes2012} and the enhancement of the low-frequency
dynamics could be associated with the pinning of those
fluctuations by impurities. These dynamics do not seem to be
involved in the superconducting mechanism  since they survive well
above the critical threshold $x_c$ for the suppression of $T_c$.

\section{\label{sec:conclusions}Conclusions}

We have shown that Y for La substitution in the optimally electron
doped LaFe$_{1-x}$Mn$_x$AsO$_{0.89}$F$_{0.11}$ superconductor
leads to a shift in the QCP driven by Mn to doping levels much
higher than the ones detected in the series without Y. This shift
is associated with an increase in the chemical pressure which
causes a decrease in the electronic correlations by Y doping,
namely in the ratio $U/t$. Both in the LaY0 and LaY20 series Mn is
observed to enhance low-frequency fluctuations in the MHz range
which are already present in the normal phase of the Mn and Y free
superconductor. These low energy fluctuations are signaled by a
peak in $1/T_1$ which is observed in different families of
iron-based superconductors and whose origin still has to be
clarified.

\section*{acknowledgments}
Brian Andersen and Maria Gastiasoro are thanked for useful
discussions. This work was supported by MIUR-PRIN2012 Project No.
2012X3YFZ2. This work has been supported by the Deutsche
Forschungsgemeinschaft through the Priority Programme SPP1458
(Grant No. BE1749/13), SFB 1143, under grant DFG-GRK1621, and
through the Emmy Noether Programme WU595/3-1 (S.W.).

\end{document}